\documentclass[reprint,superscriptaddress,amsmath,amssymb,prl]{revtex4-1}
\usepackage{graphicx}
\usepackage{dcolumn}
\usepackage{bm}
\usepackage{hyperref}
\hypersetup{colorlinks, citecolor=blue, linkcolor=blue, urlcolor=blue}
\usepackage[dvipsnames]{xcolor}
\usepackage{geometry}
\usepackage{comment}
\usepackage[skip=0pt]{caption}

\definecolor{deepgreen}{rgb}{0.1,0.5,0.1}

\definecolor{deepblue}{rgb}{0.2,0.2,0.8}

\definecolor{deepred}{rgb}{0.8,0.2,0.2}

\begin{document}
\title{Cosmological Consequences of Unconstrained Gravity and Electromagnetism}
\author{Loris Del Grosso}
\thanks{\href{mailto:loris.delgrosso@uniroma1.it}{loris.delgrosso@uniroma1.it}}
\affiliation{Dipartimento di Fisica, Sapienza Universit\`{a} di Roma, Piazzale Aldo Moro 5, 00185, Roma, Italy}
\affiliation{INFN, Sezione di Roma, Piazzale Aldo Moro 2, 00185, Roma, Italy}
\author{David~E.~Kaplan}
\thanks{\href{mailto:david.kaplan@jhu.edu}{david.kaplan@jhu.edu }}
\affiliation{The William H. Miller III Department of Physics and Astronomy, The Johns Hopkins University, Baltimore, Maryland 21218, USA}
\author{Tom Melia}
\thanks{\href{mailto:tom.melia@ipmu.jp}{tom.melia@ipmu.jp}}
\affiliation{Kavli IPMU (WPI), UTIAS, The University of Tokyo, Kashiwa, Chiba 277-8583, Japan}
\author{Vivian Poulin}
\thanks{\href{mailto:vivian.poulin@umontpellier.fr}{vivian.poulin@umontpellier.fr}}
\affiliation{Laboratoire Univers \& Particules de Montpellier (LUPM), CNRS \& Universit\'{e} de Montpellier (UMR-5299), Place Eug\`{e}ne Bataillon, F-34095 Montpellier Cedex 05, France}
\author{Surjeet~Rajendran}
\thanks{\href{mailto:srajend4@jhu.edu}{srajend4@jhu.edu}}
\affiliation{The William H. Miller III Department of Physics and Astronomy, The Johns Hopkins University, Baltimore, Maryland 21218, USA}
\author{Tristan L.~Smith}
\thanks{\href{mailto:tsmith2@swarthmore.edu}{tsmith2@swarthmore.edu}}
\affiliation{Department of Physics and Astronomy, Swarthmore College, 500 College Ave., Swarthmore, PA 19081, USA}

\begin{abstract}
Motivated by the quantum description of gauge theories, we study the cosmological effects of relaxing the Hamiltonian and momentum constraints in general relativity and Gauss’ law in electromagnetism.  We show that the unconstrained theories have new source terms that mimic a pressureless dust and a charge density that only follows geodesics. The source terms may be the simplest explanation for dark matter and generically predict a charged component. We comment that discovery of such terms would rule out inflation and be a direct probe of the initial conditions of the universe.

\end{abstract}

\maketitle

\noindent {\it Introduction}  From a distance, matter signals its existence in two ways:  via the emission of waves such as visible light and via the sourcing of a static field such as the Coulomb potential.  In the underlying classical field theories defined by a Hamiltonian, waves are a prediction of the Hamiltonian dynamics, while the static components arise from additional constraint equations imposed on the fields.  In quantum field theory, the classical dynamical equations emerge from the Schr\"{o}dinger equation in expectation value.  Conventionally, the constraints are imposed by requiring all constraint operators to annihilate physical states, effectively projecting onto a subspace of the full Hilbert space. 

In gravity, the standard constraint operators are the Hamiltonian and momentum operators.  In electromagnetism, the constraint operator is the Gauss' law operator.  As some of us have discussed \cite{Burns:2022fzs,Kaplan:2023wyw,Kaplan:2023fbl}, the quantum field theories of gravity and electromagnetism can be constructed without imposing these constraints on the initial states (for a recent exploration of the history of the subject, see \cite{Casadio:2024bdb}).  The initial state can be chosen such that the constraint operators vanish in expectation value, reproducing standard classical equations.  More generally, however, these expectation values can be non-zero and the classical equations are modified.  This formulation produces interesting consequences in electromagnetism, and is profoundly important in gravity as it is arguably the simplest solution \cite{Burns:2022fzs,Kaplan:2023wyw} to ‘the problem of time’ in quantized general relativity \cite{Page:1983uc,Unruh:1989db}.

In this article we will discuss the cosmological impact of not satisfying these constraints — Hamiltonian and momentum constraints in gravity and Gauss' law in electromagnetism.  Their violations appear as source terms in these theories and their implications are dramatic. Throughout we focus on the phenomenology that follows from assuming the standard ansatz of big-bang cosmology -- initial conditions in the form of small inhomogeneous perturbations around a mostly homogeneous initial state.  

In the case of gravity, the additional source terms behave as if there exists a pressureless dust with additional heat-flux terms that quickly redshift away outside the horizon.  This mimics a `shadow' fluid component in Einstein's equations that behaves as cold dark matter not just at the linear level in cosmological perturbation theory \cite{Kaplan:2023wyw} but, as we show here, to all orders.  Thus, for an approximately flat perturbation spectrum, we expect it to clump and virialize and to behave exactly as cold dark matter.  In the case of electromagnetism, we show that the effects are that of an inhomogeneous `shadow' charge density that follows geodesics and that does not respond to electromagnetic forces.  We expect such a background to impact the baryon-photon fluid in the early universe, structure formation in the late universe, and have interesting halo dynamics today.  

Importantly, the impact of these new source terms becomes negligible in an inflationary universe as they would redshift away during this epoch. That is, the standard classical equations of general relativity and electromagnetism are attractor solutions in an inflationary universe.  Thus, if these source terms were to explain the existence of dark matter, one would require an alternative theory in order to explain the homogeneity and initial perturbations of the universe.

\noindent {\it New Gravitational Structure as Dark Matter}
The Hamiltonian or ADM formalism of general relativity \cite{DeWitt:1967yk} is a natural context in which to quantize gravity.  Of the ten Einstein equations, six derive from the Hamiltonian equations of motion and four are imposed constraints on the fields -- the Hamiltonian constraint and the three momentum constraints. In synchronous gauge (where the time-time and time-space components of the metric $g_{\mu\nu}$ are $-1$ and 0 respectively), the constraints are equivalent to the time-time and time-space Einstein equations $\sqrt{-g}\;G^{0\mu} = 8\pi G_N \sqrt{-g}\;T^{0\mu}$ in the Lagrangian formulation. Here $G^{\mu\nu}$ is Einstein's tensor, $T^{\mu\nu}$ is the energy-momentum tensor for all other fields, $g$ is the determinant of the metric $g_{\mu\nu}$, and $G_N$ is Newton's gravitational constant.

When attempting to motivate the classical equations as a limit of the quantum field theory, the imposed Hamiltonian constraint that physical states are annihilated by the Hamiltonian, $H| \Psi\rangle = 0$, presents a significant `problem of time'.  The Schr\"{o}dinger equation in this case is now trivial and dynamical classical states (coherent states) cannot be constructed.  There have been attempts to adapt these rules to allow for some version of time evolution, often depending on the asymptotics of the state~\cite{DeWitt:1967yk} ({\it e.g.}, see \cite{Bojowald:2015iga} for a recent review of these issues in quantum cosmology).

A simple solution to the problem of time, advocated in \cite{Burns:2022fzs,Kaplan:2023wyw}, is to not impose the constraints on the states and allow for the full implied Hilbert space. One could then choose to study gravitational systems where the constraints are still satisfied at the expectation value level, {\it e.g.} $\langle \Psi| H | \Psi\rangle = 0$.  This reproduces the standard classical constraint without imposing the condition that the system is in an eigenstate of $H$, thus allowing its evolution as per the normal notion of time evolution in quantum mechanics. More generally, however, one could study systems where the constraints are not satisfied even in expectation value \footnote{Analogs were studied in Euclidean gravity\cite{Cotler:2020lxj, Cotler:2021cqa} and in Horava gravity \cite{Mukohyama:2009mz}}.  The most general classical limit  of  Einstein's equations is, in synchronous gauge
\begin{equation}\label{eq:EinsteinAux}
    \sqrt{-g}\;G^{\mu\nu} = 8\pi G_N \sqrt{-g}\,(T^{\mu\nu} + T_{\rm aux}^{\mu\nu}) ,
\end{equation} 
where
\begin{equation}\label{eq:AuxT}
   \sqrt{-g}\; T_{\rm aux}^{\mu\nu} \equiv 
    \begin{pmatrix}
    \mathbb{H} & \mathbb{P}^1 & \mathbb{P}^2 & \mathbb{P}^3 \\
    \mathbb{P}^1 & 0 & 0 & 0 \\
    \mathbb{P}^2 & 0 & 0 & 0 \\
    \mathbb{P}^3 & 0 & 0 & 0 
    \end{pmatrix}
\end{equation}
and where $\mathbb{H}$ and $\mathbb{P}^i$ are functions of spacetime. The Bianchi identity $\nabla_\mu G^{\mu\nu} = 0$ and a covariantly conserved energy-momentum tensor $\nabla_\mu T^{\mu\nu} = 0$, as a result of the equations of motion of the other fields, imply that this auxiliary tensor is also covariantly conserved, setting the following conditions in synchronous gauge:
\begin{eqnarray}
\partial_0 \mathbb{H} = - \partial_i \mathbb{P}^i \label{eq:SHcontinuity}\\
\partial_0 \left(g_{ij}\mathbb{P}^j \right) = 0 .\label{eq:SHeuler}
\end{eqnarray}
We now study the physical consequences of these additional source terms.

Let us first analyze the case where  the $\mathbb{P}^i=0$.
Then, from Eq.~\eqref{eq:SHcontinuity}, we see that $\mathbb{H}=\mathbb{H}({\bf x})$ is independent of time. 
If $\mathbb{H}$ is everywhere positive, we can additionally remove the remaining coordinate freedom in the synchronous gauge and make a spatial coordinate redefinition ${\bf x}\rightarrow {\boldsymbol{\xi}}({\bf x})$ such that the determinant $||\partial x^i/\partial \xi^j|| = \overline{\mathbb{H}}/\mathbb{H}({\bf x})$, where $\overline{\mathbb{H}}$ is a constant.  Thus, in this case, the spatial dependence in $\mathbb{H}$ can be entirely moved into the metric. We will briefly discuss non-positive definite $\mathbb{H}$ later in the article.

We can show that this new source term will reproduce the effects of dark matter by noting that our auxiliary energy-momentum tensor can be written as
\begin{equation}\label{eq:DarkMatter}
    T_{\rm aux}^{\mu\nu} = \rho_{\rm aux} u^\mu u^\nu
\end{equation}
in which $\rho_{\rm aux} = \overline{\mathbb{H}}/\sqrt{-g}$ and $u^\mu = \{1,0,0,0\}$ in synchronous gauge.  The energy momentum tensor \eqref{eq:DarkMatter} is recognizably the form of a general pressureless dust, $\nobreak{T^{\mu\nu}=\rho u^\mu u^\nu}$. In synchronous gauge, a dust with zero velocity will satisfy
\begin{eqnarray*}
    0 &=& \nabla_\mu T^{\mu\nu} \\
    &=& \partial_\mu (\rho u^\mu u^\nu ) + \Gamma_{\mu\lambda}^\mu \rho u^\lambda u^\nu + \Gamma_{\mu\lambda}^\nu \rho u^\mu u^\lambda \\
    &=& \delta_0^\nu \partial_0 \rho + (1/\sqrt{-g}) (\partial_0 \sqrt{-g}) \delta_0^\nu \rho \\
    &=& (1/\sqrt{-g}) \partial_0 (\sqrt{-g} \rho)\delta_0^\nu
\end{eqnarray*}
where in the third line we used the relation $\partial_\lambda \sqrt{-g} = \Gamma^\mu_{\mu\lambda}\sqrt{-g}$ and the fact that $\Gamma_{\mu0}^0 = 0$ in synchronous gauge. The above implies $\rho = f({\bf x})/\sqrt{-g}$ is the most general form of pressureless, zero-velocity dust in synchronous gauge (where again we could choose to absorb the ${\bf x}$ dependence into the metric). That is,  $\rho_{\rm aux}$ behaves precisely like pressureless dust.

To reinforce the above point, consider the following equation
\begin{equation}\label{eq:Tmunuconservation}
    0 = \nabla_\mu  T_{\rm aux}^{\mu\nu} = \nabla_\mu (\rho_{\rm aux}\, u^\mu) u^\nu + \rho_{\rm aux} \, u^\mu \nabla_\mu u^\nu \, .
\end{equation}
In synchronous gauge it is particularly easy to see that 
 the four-velocity $u^\mu$ satisfies the geodesic equation, $\nobreak{u^\mu\nabla_\mu u^\nu = 0}$, and thus the density current $\rho_{\rm aux}u^\mu$ is covariantly conserved
\begin{equation}\label{eq:conservation_energy_density}
    \nabla_\mu (\rho_{\rm aux}\, u^\mu) = 0 \,.
\end{equation}
Being tensor equations, these are true in any gauge.  Thus, this system will evolve as if there is a fluid component of pressureless dust that undergoes geodesic evolution. Assuming our ansatz of nearly homogeneous initial conditions, we predict evolution equivalent to that of cold dark matter.

Since simulations of dark matter implicitly assume the dynamics are independent of the short-wavelength physics  (in fact it is well known that short-wavelength nonlinearity quickly decouples from long-wavelength evolution \cite{Baumann:2010tm}), the usual machinery of $N-$body simulations or perturbation theory \cite{Bernardeau:2001qr} can be used to study the non-linear evolution here as well. Similar to the case of real dark matter, to avoid issues of potential coordinate singularities in the synchronous gauge, one can transform the equations to another gauge such as Newtonian gauge  \cite{Ma:1995ey}. In the quantum theory,  this amounts to identifying a basis of coherent states that smoothly tracks the classical evolution. In this way, for the case of both shadow and real dark matter, a system with initial conditions set in the synchronous gauge can be evolved to arbitrary times in the future.

One ultimately expects that virialized structures such as halos will form exactly as in the standard cold dark matter picture. Barring standard gravitational effects ({\it e.g.}, tidal stripping), the effect of baryons or other new physics unrelated to this type of matter, structures may form down to very small scales, potentially even below the free-streaming length of more conventional dark matter candidates such as weakly interacting massive particles.

We now proceed to consider the most general case of the auxiliary tensor where  $\mathbb{P}^i\ne0$.  The source term can be written in the form
\begin{equation}\label{eq:TmunuSMconvariant}
    T_{\rm aux}^{\mu\nu} = \rho_{\rm aux}  \,u^\mu u^\nu + q^\mu u^\nu + u^\mu q^\nu \,,
\end{equation}
where the new spatial vector $q_{\mu}$ satisfies $q_{\mu} u^\mu = 0$ and $q_\mu q^\mu \ge 0$. In synchronous gauge, we take $\rho_{\rm aux} = \mathbb{H}/\sqrt{-g}$, $u^\mu = \{1,0,0,0\}$, and $q^\mu = (0, \mathbb{P}^1, \mathbb{P}^2,\mathbb{P}^3)/\sqrt{-g}$. Note now both $\mathbb{H}$ and $\mathbb{P}^i$ are generally functions of space and time. The covariant conservation of this tensor now gives
\begin{eqnarray}\label{eq:fullTmunuconservation}
    0 &=& \nabla_\mu  T_{\rm aux}^{\mu\nu} \\ \nonumber
    &=& \nabla_\mu (\rho_{\rm aux}\, u^\mu) u^\nu + \rho_{\rm aux} \, u^\mu \nabla_\mu u^\nu \, \\ \nonumber
    && +  \left(\nabla_\mu q^\mu\right) u^\nu + q^\mu \nabla_\mu u^\nu \\ \nonumber
    && + \left(\nabla_\mu u^\mu\right) q^\nu + u^\mu \nabla_\mu q^\nu .
\end{eqnarray}
The density $\rho_{\rm aux}$ is now not covariantly conserved, 
which we can see by contracting \eqref{eq:fullTmunuconservation} with $u_\nu$:
\begin{eqnarray} \label{eq:contwithq}
     \nabla_\mu(\rho_{\rm aux} u^\mu) = - \nabla_\mu q^\mu \,,
\end{eqnarray}
where we have imposed the geodesic equation and the various identities above. 
Similarly, contracting onto the orthogonal direction to $u^\nu$ with the projector $u_\nu u_\lambda + g_{\nu\lambda}$, we have a constraint on $q^\mu$:
\begin{eqnarray} \label{eq:eulerwithq}
     0 =q^\mu  \nabla_\mu u_\lambda + \left(\nabla_\mu u^\mu\right)q_\lambda + u^\mu \nabla_\mu q_\lambda\,.
\end{eqnarray}
In synchronous gauge, eqs.~\eqref{eq:contwithq} and \eqref{eq:eulerwithq} simply reduce to eqs.~\eqref{eq:SHcontinuity} and \eqref{eq:SHeuler}, respectively. Thus we see that there is an additional effect on the auxiliary density in this case.  While the four-vector field $u^\mu$ follows geodesics by design, the $q^\mu$ acts as a `heat flux' on the density $\rho_{\rm aux}$ causing an additional evolution of the effective mass density of the fluid.


Let us now analyse the effect of this more general source term within our comological ansatz.  We write the source terms as $\mathbb{H}({x}) \equiv \overline{\mathbb{H}} + \delta\mathbb{H}({x})$  and $\mathbb{P}^i({x}) \equiv \delta\mathbb{P}^i({x})$, where the $\delta$'s represent linear-order perturbations around a homogeneous and isotropic background.  Here, taking the spatial metric at leading order to be $g_{ij} \sim a(t)^2\delta_{ij}$, we can see from the constraint equation \eqref{eq:SHeuler} that $\partial_0 (\delta \mathbb{P}^i a^2) \simeq 0$, or $\delta \mathbb{P}^i/\sqrt{-g} \sim a^{-5}$.  Thus this contribution to $T_{\rm aux}$ redshifts quickly even outside the horizon. 

While it is reasonable to consider the $\delta \mathbb{P}^i$ terms negligible at late times for a wide range of initial conditions, it is interesting that for a flat spectrum, they may in principle be important for gravitational bound states at the smallest scales. 
In addition, the consequences of non-zero  $\delta\mathbb{P}^i({\bf x},t)$ at early times  could be much more dramatic. A uniform
$\delta\mathbb{P}^i$ in the early universe would be a source of anisotropy.
The most stringent constraints should thus come from measurements of the primordial abundance of light elements and big-bang nucleosynthesis, as current measurements (except for $^{7}$Li) favor a homogeneous universe that dilutes like radiation with few percent precision at $z\in[10^8-10^{10}]$ \cite{An:2023buh,Hertzberg:2024uqy}. However, we note that most analyses focus on the predictions of the Bianchi Type I universe, which does not describe the dynamics of the model we are considering here.  For more generic $\delta\mathbb{P}^i$, estimating the constraints would require a dedicated analysis.

\noindent {\it Shadow Charges in General Relativity}  Standard electromagnetism can be formulated in the Hamiltonian language, in which case Amp\`ere's law comes from Hamilton's equations of motion.  On the other hand, Gauss' law, $\partial_\mu F^{\mu 0} - J^0 = 0$, is a constraint in flat space.  The effect of not enforcing this constraint allows for a source term that behaves as a static charge density \cite{Kaplan:2023fbl} (for some applications of the Hamiltonian formalism and loosening this constraint in electromagnetism, see \cite{Gervais:1978kn,Pisarski:2022cuo}).  In curved space, the equations of motion can similarly be supplemented by an auxiliary charge density
\begin{equation}\label{eq:Maxwell}
     \nabla_\mu F^{\mu\nu} = (J^\nu + J_{\rm aux}^\nu)
\end{equation}
where $\nabla_\mu$ is now a covariant derivative and $J^\nu$ is the four-current associated to standard matter.  Taking the divergence of Equation \eqref{eq:Maxwell}, and accounting for charge conservation among normal charged matter, we have
\begin{equation}\label{eq:ChargeConservation}
    \nabla_\mu J_{\rm aux}^\mu = 0
\end{equation}
In synchronous gauge, we can define 
\begin{equation}\label{eq:AuxCurrent}
    \sqrt{-g} J_{\rm aux}^\mu \equiv \{ \mathbb{J},0,0,0\}
\end{equation}
where \eqref{eq:ChargeConservation} requires $\mathbb{J}=\mathbb{J}({\bf x})$ to be time-independent \cite{Kaplan:2023fbl}.  The auxiliary current can thus be written as
\begin{equation}
    J_{\rm aux}^\mu = \rho_{\rm aux}^{ch} v^\mu
\end{equation}
with $\rho_{\rm aux}^{ch} = \mathbb{J}/\sqrt{-g}$ and $v^\mu=\{1,0,0,0\}$ in synchronous gauge.  Because $v^\mu\nabla_\mu v^\nu = 0$ in all coordinate systems, this auxiliary charge density follows geodesics and thus does not respond directly to electromagnetic forces.  In addition, the conservation equation \eqref{eq:ChargeConservation} is equivalent to the auxiliary matter conservation equation \eqref{eq:conservation_energy_density}.  Thus, this new fluid acts like a charged component of the auxiliary matter.

In addition, while the shadow charge itself does not represent a physical field in the universe, the electric fields it sources are and thus contribute an additional source for the metric via Einstein's equations.   The energy-momentum tensor for all electromagnetic fields and charged matter is
\begin{equation}
    T^{\mu\nu} = \mathcal{E}^{\mu\nu} + T^{\mu\nu}_{matter} \equiv F^{\mu\lambda}F^{\nu}_{\lambda} - \frac{1}{4} g^{\mu\nu} F^{\lambda\sigma}F_{\lambda\sigma} + T^{\mu\nu}_{matter}
\end{equation}
If we include the violation of Gauss' law \`{a} la \eqref{eq:Maxwell}, we find this energy-momentum tensor is not covariantly conserved 
\begin{eqnarray}
    \nabla_\mu T^{\mu\nu} &=&  \nabla_\mu \mathcal{E}^{\mu\nu} + \nabla_\mu T^{\mu\nu}_{matter}\nonumber \\
    &=& F^\nu_{{\phantom \nu}\lambda} (J^\lambda + J_{\rm aux}^\lambda) - F^\nu_{{\phantom \nu}\lambda} J^\lambda 
\end{eqnarray}
where we have used
\begin{equation}
    F^{\mu\lambda}\nabla_\mu F^\nu_\lambda - \frac{1}{2} g^{\mu\nu}F^{\lambda\sigma}F_{\lambda\sigma} = 0
\end{equation}
which can be shown using the Bianchi identity. 

Therefore, in order to guarantee the covariant conservation of the full energy-momentum tensor, we must add a new component $T_{\rm aux}^{\mu\nu}$, describing the contribution of the new auxiliary charged matter. 
Again, in synchronous gauge we parameterize $T_{\rm aux}^{\mu\nu}$ as in \eqref{eq:AuxT}, but with new constraint equations:
\begin{eqnarray}
    \partial_0 \mathbb{H} + \partial_i \mathbb{P}^i = 0\\
    \partial_0 \left(g_{kj}\mathbb{P}^j\right) = -  F_{k0} \mathbb{J}
    \label{eq:chargemattercoupling}
\end{eqnarray}
These equations can be integrated to solve explicitly for $\mathbb{H}$ and $\mathbb{P}^i$.  In an expanding universe, these new terms redshift away if they are small (perturbative) compared with the homogeneous energy density.

While the time-dependence of this auxiliary fluid is fixed by the constraints, the remaining freedom in the initial conditions is of course arbitrary.  Starting with our ansatz, we expect the auxiliary current to contribute at first order, {\it i.e.}, small inhomogeneous perturbations in the shadow charge density about zero.  Describing these fluctuations in Fourier space, the modes outside the horizon in an expanding universe will evolve such that the charge density redshifts like a matter density, as the inverse of the scale factor cubed.  

Inside the horizon, shadow charge is a source of rich phenomenological possibilities with  complex dynamics.  Because shadow charge flows with the dark matter but couples to the baryon-photon fluid, it could represent an effective interaction between dark matter and baryonic matter.  This interaction could show up in the early universe affecting the photon diffusion length and the ionized matter fraction after recombination.  It could be a source of isocurvature perturbations if the perturbations were distinct from shadow matter.  And it could also be a new source of signals in the galaxy today.  For example, the shadow charge could be screened by regular matter, but this matter could interact at short distances with the Earth's atmosphere or with cosmic ray experiments on the Earth's surface. It may also lead to a new explanation for the large-scale magnetic fields observed in galaxy clusters and cosmic filaments \cite{Kronberg:1993vk,Carretti:2022tbj,Vernstrom:2021hru}. The complexity of the interaction of  shadow matter with regular matter would be an exciting source of new signals and motivates numerical simulations on cosmological and galactic scales.

{\it Early universe cosmology} An important point about this new modification to Einstein's equations is that $T^{\mu\nu}_{\rm aux}$ is not dynamical, in the sense that it is not adding additional degrees of freedom with wave-like properties. The new source term parameterizes the initial conditions of the non-tensor parts of the gravitational field and thus represents a kind of relic structure.  Due to the redshift properties of this  source, a period of standard cosmological inflation (for a review, see \cite{Baumann:2009ds}) would completely dilute this contribution to cosmology, in which case it could not play the role of dark matter.  If shadow matter was in fact the dark matter, it would point to a different early cosmology and would be a direct probe of the initial conditions of the universe.

In light of the above, it is of interest to explore the evolution of shadow matter and charge in non-inflationary, early universe cosmologies, such those with quasi-static or oscillating periods which decay into an expanding universe \cite{Wands:1998yp,Steinhardt:1999nw,Steinhardt:2001st,Finelli:2001sr,Graham:2011nb,Horn:2017kmv, Graham:2017hfr}.  Note that if the matter and radiation in the universe have their own set of perturbations, this extra inhomogeneity would represent a contribution to isocurvature. These uncorrelated dark matter density isocurvature perturbations would be constrained by cosmological observations such as Planck data \cite{Planck:2018jri}. 
Thus it would be interesting to see if one can be dominated by adiabatic perturbations in these other scenarios.

Critically, all of the above alternatives to inflation require violations of the null-energy condition (NEC).  For a fluid described by the stress-energy tensor in Eq.~\eqref{eq:TmunuSMconvariant} the NEC is trivially violated by choosing $\mathbb{H} < 0$. However, there is also the possibility of having positive $\mathbb{H}$, while requiring
\begin{equation}\label{eq:energy_condition_one}
    \rho_{\rm aux}^2 < 4 Q^2 \, ,
\end{equation}
where $Q^2 \equiv q_\mu q^\mu$. This is enough to violate all the energy conditions (including the NEC)~\cite{Maeda:2022vld}. Using the solution of Einstein's equation to linear order around a homogeneous and isotropic universe, the inequality~\eqref{eq:energy_condition_one} can be met at some small but finite value of the scale factor, assuming the remaining matter dominates and is in the linear regime.  Upon further contraction, if the NEC-violating source terms become comparable to the background quantities, the perturbative approximation breaks down. These effects strongly motivate studying the full nonlinear regime, where this possibility of violating the NEC may give rise to interesting effects beyond the non-singular bounces described above, such as wormhole geometries ({\it e.g.}, \cite{Morris:1988tu,Maldacena:2018gjk}), structures with negative mass, and other phenomena.

{\it Discussion} In this article we showed how modifications of the source terms in general relativity generate a `shadow matter' that could be the perceived dark matter in the universe, and how modifications of Gauss' law generate a `shadow charge'  contribution to the Maxwell-Einstein equations.  A shadow matter explanation of dark matter can be immediately falsified by an experimental discovery of inflation. Any positive results from particle dark matter direct or indirect detection experiments would also rule out shadow matter from being (all of) the dark matter. However, assuming such evidence does not materialize, there are  avenues that can potentially distinguish shadow matter from dynamical dust.  For example, the heat flux terms in the auxiliary stress-energy tensor that may be small in the present universe could have had an observable effect at early times.  More exotically, small regions could exist where the shadow matter has negative energy density. Either of these would be a smoking gun of a shadow matter fluid.

But perhaps the most promising avenue to search for evidence of shadow matter is to pursue experimental efforts to observe shadow charge. Observation of shadow charge would, by itself, rule out cosmic inflation, demonstrate that the universe violates Gauss' law, and thus render the existence of generic shadow matter extremely plausible. More than this, it would in fact be a direct verification of the existence of at least some shadow matter component to the universe, by the fact that shadow charge sources its own shadow matter fluid.

One might wonder whether both the shadow matter and shadow charge can be captured simply as decoupling limits of theories with dynamical fields -- the former being matter with all interactions (except gravity) set to zero, while the latter being charged matter in the infinite mass limit.  However, the new terms for the gravity equations can be sourced by null-energy violating shadow matter (for example, $\mathbb{H}$ could be negative in some regions).  And of course the infinite mass limit to mimic shadow charge is not viable when coupled to gravity.  

Thus, these  source terms represent a new consistent phenomenological effect that would not have been explored in any known part of parameter space of new dynamical degrees of freedom. Indeed, we emphasize that they do not stem from any modifications to the Lagrangian used to form the path integral of the quantum theory. Rather, they should be viewed simply as arising from the freedom to set the initial conditions needed to fully specify the path integral as a solution to the Schr\"odinger equation.  Initial conditions do not represent new dynamical degrees of freedom, but of course they can influence the ensuing dynamical evolution, in precisely the way we have described.

The evidence for dark matter could be a signal that the constraint equations of general relativity and electromagnetism are violated in our universe.  If so, a rich `shadow world' phenomenology could be expected, the observation of which would directly probe the initial state and evolution of the universe.

\noindent {\it Acknowledgements} We thank  Raman Sundrum for many enlightening conversations.  We also thank Daniel Grin, Marc Kamionkowski, Alvise Racanelli, Alfredo Urbano, and Neal Weiner for useful discussions.

This work was supported by the U.S.~Department of Energy~(DOE), Office of Science, National Quantum Information Science Research Centers, Superconducting Quantum Materials and Systems Center~(SQMS) under Contract No.~DE-AC02-07CH11359. 
TM is supported by the World Premier International Research Center Initiative (WPI) MEXT, Japan, and by JSPS KAKENHI grants JP19H05810, JP20H01896, JP20H00153, JP22K18712, and JP24H00244. 
DEK and SR is supported in part by the U.S. National Science Foundation (NSF) under Grant
No. PHY-1818899. 
SR is also supported by the Simons Investigator Grant No. 827042,
and by the DOE under a QuantISED grant for MAGIS. 
DEK is also supported by the Simons Investigator Grant No. 144924.
LDG acknowledges the Johns Hopkins
University for hospitality during the completion of this
project and H2020-MSCA-RISE-2020 GRU (Grant agreement ID: 101007855) for financial support. 
VP has received support from the European Union’s Horizon 2020 research and innovation program under the Marie Skodowska-Curie grant agreement No 860881-HIDDeN. VP has received funding from the European Research Council (ERC) under the
European Union’s HORIZON-ERC-2022 (Grant agreement No. 101076865). 
TLS is supported by NSF Grants No.~2009377 and No.~2308173. TLS and DEK thank the NYU CCPP where part of this work was completed.

\bibliography{biblio}
\end{document}